\documentclass[useAMS,usenatbib, usegraphicx, manuscript]{emulateapj}
\pdfoutput=1
\usepackage{fancyhdr}
\usepackage{hyperref}
\usepackage{wasysym}
\usepackage{longtable}
\usepackage{aas_macros}
\usepackage{txfonts}
\usepackage{times}
\usepackage{graphicx}
\usepackage{color}
\usepackage{ulem}
\input{astro.sty}

\usepackage[T1]{fontenc}% (2) specify encoding

\shorttitle{Characterizing crosstalk within the Pan-STARRS GPC1 camera}
\shortauthors{T.J.L. de Boer et al}
\begin{document}
\title{Characterizing crosstalk within the Pan-STARRS GPC1 camera}
% this is a crude trick to get the order of affiliations right.  These
% names are used in the affiliations below.  The user needs to (1) set
% the order and numbers to have the correct sequence in the author
% list and (2) re-order the list at the bottom (and comment-out as needed)
\def\IfA{1}
%\def\CfA{2}
%\def\LBL{3}
%\def\Hubble{4}
%\def\DUH{5}

% This example has a first author from UH:
\author{
T. J.L. de Boer,\altaffilmark{\IfA}
M. E. Huber,\altaffilmark{\IfA}
E. A. Magnier,\altaffilmark{\IfA}
P. M. Onaka,\altaffilmark{\IfA}
K.~C. Chambers,\altaffilmark{\IfA} 
C.-C. Lin,\altaffilmark{\IfA} 
H. Gao,\altaffilmark{\IfA} 
J. Fairlamb,\altaffilmark{\IfA} and
R. J. Wainscoat\altaffilmark{\IfA} 
%J.~L. Tonry, \altaffilmark{\IfA}
%D. Finkbeiner,\altaffilmark{\CfA}
%E. Schlafly,\altaffilmark{\LBL,\Hubble}
%PS Builder List
%W.~S. Burgett,\altaffilmark{\IfA}
% L. Denneau,\altaffilmark{\IfA}
% P. Draper,\altaffilmark{\DUR}
%H.~A. Flewelling,\altaffilmark{\IfA}
% T. Grav,\altaffilmark{\IfA}
% J. N. Heasley,\altaffilmark{\IfA}
%K. W. Hodapp,\altaffilmark{\IfA}
% R. Jedicke,\altaffilmark{\IfA}
%N. Kaiser,\altaffilmark{\IfA}
%R.-P. Kudritzki,\altaffilmark{\IfA}
% G. A. Luppino,\altaffilmark{\IfA}
% R. H. Lupton,\altaffilmark{\Princeton}
% E. A. Magnier,\altaffilmark{\IfA}
%N. Metcalfe,\altaffilmark{\DUH}
% D. G. Monet,\altaffilmark{\USNO}
% J.~S. Morgan,\altaffilmark{\IfA}
% P.~A. Price,\altaffilmark{\Princeton}
% C.~W. Stubbs,\altaffilmark{\CfA}
% W.~E. Sweeney,\altaffilmark{\IfA}
% J.~L. Tonry, \altaffilmark{\IfA}
%R. J. Wainscoat,\altaffilmark{\IfA} and 
%C. Z. Waters,\altaffilmark{\IfA}
} % this bracket terminates author list

% The ordering here should be sequential, matching the sequence in the list of authors:
\altaffiltext{\IfA}{Institute for Astronomy, University of Hawaii, 2680 Woodlawn Drive, Honolulu HI 96822}
%\altaffiltext{\CfA}{Harvard-Smithsonian Center for Astrophysics, 60 Garden Street, Cambridge, MA 02138}
% \altaffiltext{\Princeton}{Department of Astrophysical Sciences, Princeton University, Princeton, NJ 08544, USA}
% \altaffiltext{\USNO}{US Naval Observatory, Flagstaff Station, Flagstaff, AZ 86001, USA}
% \altaffiltext{\JHU}{Department of Physics and Astronomy, Johns Hopkins University, 3400 North Charles Street, Baltimore, MD 21218, USA}
% \altaffiltext{\MPIA}{Max Planck Institute for Astronomy, K\"onigstuhl 17, D-69117 Heidelberg, Germany}
%\altaffiltext{\DUH}{Department of Physics, Durham University, South Road, Durham DH1 3LE, UK}
%\altaffiltext{\LBL}{Lawrence Berkeley National Laboratory, One Cyclotron Road, Berkeley, CA 94720, USA}
%w\altaffiltext{\Hubble}{Hubble Fellow}

\begin{abstract}
Using data from a year-long dedicated campaign to observe bright stars, we study the crosstalk channels present in the GPC1 camera. By analyzing these data, we construct a dataset that checks source stars on almost every CCD of every chip within the camera against all possible crosstalk destinations.

We use a clustering algorithm to find potential crosstalk occurrences, and then also check all possible combinations (driven by the hardware layout) by eye. This results in a total of 640 rules, with a flux attenuation factor ranging from 2.5$\times$10$^{2}$ for the bright end to 2.5$\times$10$^{4}$ at the faint end. The average value of m$_{cross}$-m$_{src}\approx$-10.25 corresponds to an attenuating factor of 1.25$\times$10$^{4}$, which produces crosstalk ghosts with an average signal-to-noise ratio of 0.64$\pm$0.1 on the bright images. We find no evidence of crosstalk signals between CCDs not connected in the hardware setup. 

The distribution of attenuation factors is also found to be dependent on crosstalk movement. A clear dependence on cell column offsets is found, consistent with the idea that the source star charge is progressively attenuated during the traversal of cell readout lines. While we can see the trends, the uncertainties on aperture magnitude measurements are large at this stage. 
\end{abstract}

% insert additional keywords as appropriate:

\section{INTRODUCTION}\label{sec:intro}
The Panoramic Survey Telescope and Rapid Response System (Pan-STARRS) is a wide-field astronomical imaging and data processing facility developed and operated at the University of Hawaii’s Institute for Astronomy \citep{Kaiser2002, Kaiser2010}. The Pan-STARRS facility consists of two 1.8m telescopes located on the summit of Haleakala on the Hawaiian island of Maui. The Pan-STARRS1 (PS1 for short) telescope began formal operations on May 13 of 2010, while the Pan-STARRS2 (PS2 for short) telescope is currently being commissioned \citep{Chambers2016}.

The first large-scale public data release of Pan-STARRS (DR1) contained the results of the Pan-STARRS 3$\pi$ Survey, and was released on 16 December 2016. This release contained only average information resulting from the many individual images obtained by the 3$\pi$  Survey observations. A second data release, DR2, provides measurements from all of the individual exposures, and includes an improved calibration of the PV3 processing of the DR1 dataset, and was made available 28 January 2019.

The Pan-STARRS data handling is primarily done by the Image Processing Pipeline (IPP), which consists of a suite of software programs and data systems that reduce the observed images, measure astronomical sources, perform calibrations \citep{Magnier2020,Magnier2020_2,Magnier2020_3}. The processing system is set up in such a way as to process the large amount of data generated by the PS1 telescope during the night they are observed and therefore includes extensive parallelization across a large cluster of computers. Following the data reduction and calibration, the produced products are distributed to the various user communities, for further use in conducting scientific analyses, object characterization and follow-up \citep{Flewelling2020}.

The primary science design drivers for PS1 included a number of science goals focused among others on studying the Milky Way, M31 and our own Solar System using a series of dedicated surveys \citep{Chambers2016}. Of particular importance here is the goal of Pan-STARRS to survey our Solar System for Potentially Hazardous Objects among Near Earth Asteroids \citep{Chambers2016}. The Pan-STARRS data is used for the detection of both moving objects (e.g., asteroids) and variable source such as explosive transient sources (e.g., supernovae). To facilitate the detection of moving or variable objects, data is during the night in so-called chunks where an area of the sky is revisited  several (typically 4) times. This allows us to construct difference images, which are commonly used to remove the clutter of static stars and galaxies. Within the Pan-STARRS system, difference images are generated using the PSF-matching technique described by \citet{Alard1998} before being passed on to the Pan-STARRS Moving Object Processing System (MOPS; \citet{Denneau2013}) for analysis. 

Any object that remains within on difference images is a potential moving or variable object, and will be looked at by eye for evaluation, if it passes further checks on data quality and orbit analysis. Therefore, it is crucial to remove or flag any potential contaminants that can pose as a moving object. Given that the difference images are constructed from multiple visits to the same location while the sky moves overhead in the intervening time interval, instrumental artifacts and optical features (such as ghosts and glints) must be carefully tracked and masked. Most of these sources of contamination are detected and characterized within the IPP system during the Camera stage \citep[see][for details]{Magnier2020}. 

% EAM : I re-wrote much of this paragraph in an attempt to 
% make it clear that the crosstalk images are generated in the signal chain, not the detector itself.
Among those features, crosstalk is a source of contamination that is difficult to find and characterize. Crosstalk is the appearance of an electronic ``ghost'' object in the output data from one detector generated by the signal of a bright star elsewhere within the focal plane. The signal of the bright star is transferred electrically between CCD readout chains during the readout process, typically between amplifiers or readout lines during simultaneous clocking operations.  The phenomenon should not be confused with persistence, which also creates ghost objects in the imaging data.   Persistence is due to the charge persisting within the silicon of the CCD and being picked up during a subsequent readout.   
%This could be through cables running close together, or through charge of multiple CCDs being relayed through the same controller.

In this paper, we examine the presence and behavior of crosstalk in the PS1 Gigapixel Camera 1 (known as GPC1) CCDs. While the presence of crosstalk has been investigated previously as part of the PS1 commissioning process, no study of the crosstalk behavior within the actively running Pan-STARRS system has been conducted. This work will describe and characterize the crosstalk rules currently in effect within the PS1 system, and be used as a blueprint for the identification and masking of the crosstalk ghost pixels within the existing IPP framework.

In Section \ref{sec:method}, we will discuss the PS1 camera in more detail, as well as the dedicated campaign to obtain data suitable for this study. The technique and methods employed to extract the crosstalk rules is also discussed. In Section \ref{sec:results}, we present the set of rules found within the PS1 system, characterize the crosstalk attenuation factors and show the behavior as function of hardware layout.  Finally, we conclude in section \ref{sec:conclusions} with a discussion of the implications of this work for future facilities and moving object detection in general. 
\begin{figure}
\centering
\includegraphics[angle=0, width=0.495\textwidth]{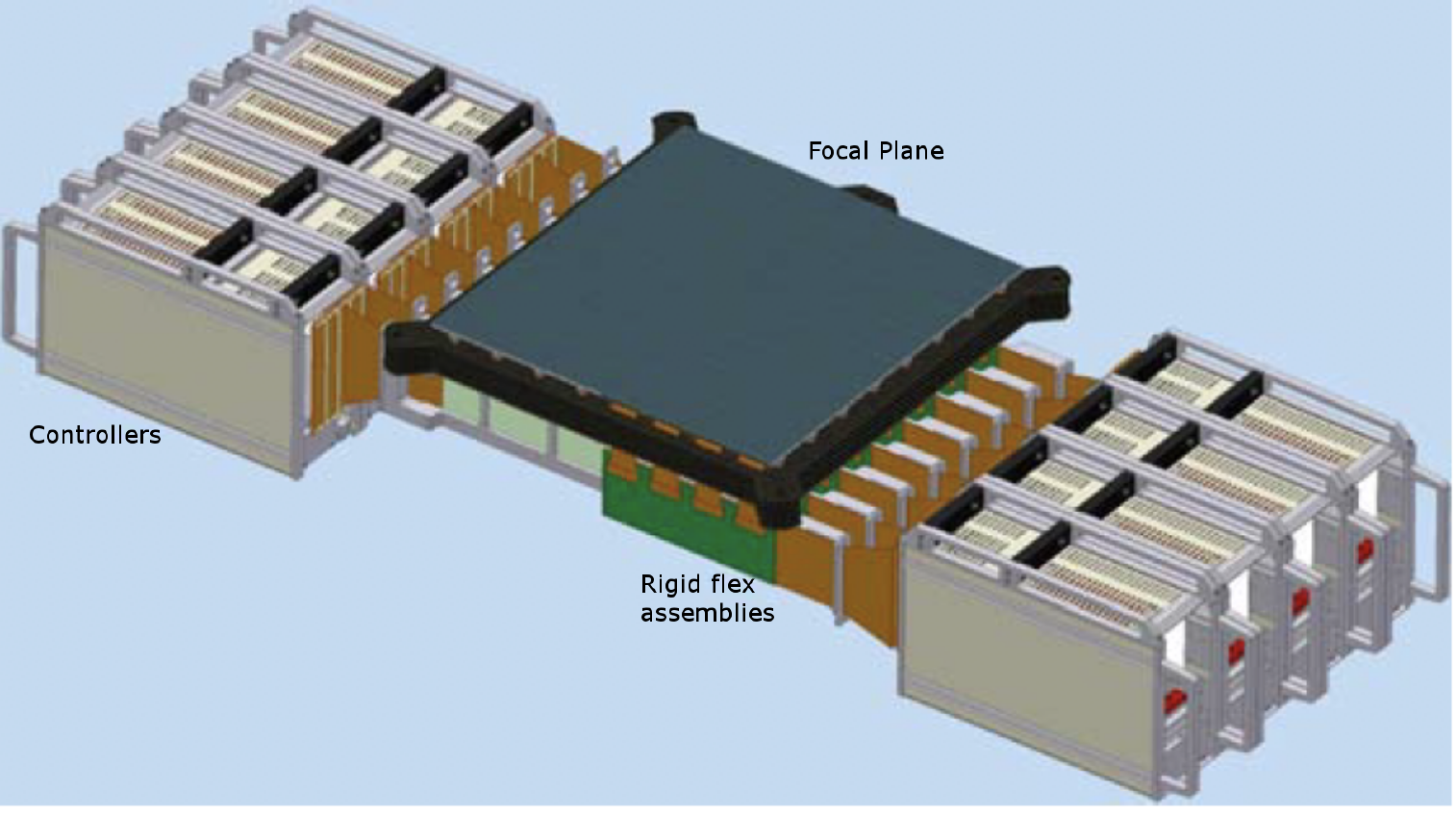}
\caption{The layout of STARGRASP controllers relative to the focal plane within the Pan-STARRS system.} \label{fig:OTA_form_factor}
\end{figure}

\section{METHOD}\label{sec:method}
Before discussing the methodology behind finding the different crosstalk channels, it is important to understand the nomenclature and layout of the telescope and camera. The wide-field optical design of the PS1 telescope \citep{Hodapp2004} produces a 7 square degrees field of view with low distortion and minimal vignetting even at the edges of the illuminated region. To achieve this, GPC1 consists of a mosaic of 60 densely packed CCD Orthogonal Transfer Arrays (OTAs) manufactured by Lincoln Laboratory \citep{Tonry2009}. They are laid out in an 8$\times$8 grid, from which the corners are missing. These OTAs themselves each consists of an 8$\times$8 grid of 590$\times$598 pixel readout regions (referred to as cells), yielding an effective 4846$\times$4868 detector. Initial performance assessments are presented in \citet{Onaka2008}.

\begin{figure*}
\centering
\includegraphics[angle=0, width=0.95\textwidth]{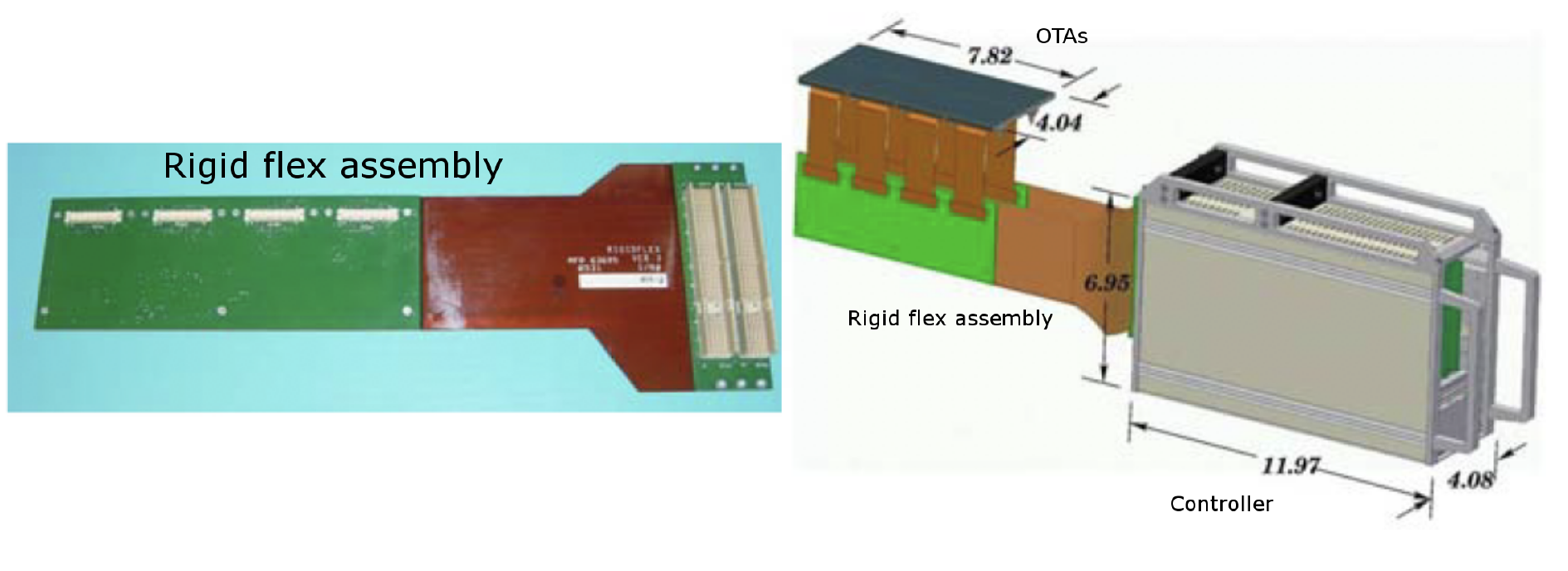}
\caption{Pan-STARRS rigid-flex assembly with slots for four OTAs (left) and one of the STARGRASP controllers housing two rigid-flex assemblies (right).} \label{fig:stargrasp_rigidflex}
\end{figure*}

Driven by the short cassegrain depth of the telescope and the size of the focal plane, the GPC1 physical form factor was laid out in such a way that the controllers were placed outside the focal plane mosaic (see Fig. \ref{fig:OTA_form_factor}). Given the layout, the STARGRASP controllers on one side of the focal plane are 'upside down' (rotated by 180\degrees) with respect to the other side of the focal plane, and native pixel coordinates therefore run in opposite directions. Each STARGRASP controller itself is connected to a rigid-flex assembly (a combination of a flex cable and printed circuit board) which penetrates the dewar wall and houses slots for 4 OTAs (see Fig.~\ref{fig:stargrasp_rigidflex}). Two sets of rigid-flexes are connected to a chassis housing controller electronics \citep{Onaka2012}.

Given the setup described above, the important facts regarding the focal plane and readout of the GPC1 camera can be described as follows:
\begin{itemize}
    \item The focal plane is made up of 60 OTAs numbered according to their column and row within the layout, counted from the bottom right to the left and up, starting from zero (i.e. OTA XY20 represents the OTA in the third column from the right, in the first row).
    \item Each OTA consists of 64 cells, which are similarly counted from the bottom right to the left and up, \textbf{within the reference frame of the controller}. Seen from the focal plane view, this means cell xy43 on OTA XY20 is the fifth column, fourth row counting from the bottom right and cell xy43 on OTA XY40 is the fourth column, fifth row counting from the bottom right.
    \item All of the OTA rows are read out together.
    \item Within a flex print, alternate OTAs are read out simultaneously through the controllers. This means that e.g. OTA XY40 and OTA XY42 are read out simultaneously, but at a different time from OTA XY41 and OTA XY43, which are also read out together at the same time.
    \item Within a single OTA, an entire row of cells is read out simultaneously. This means cells xy00-xy07 are read out at the same time, followed by xy10-xy17, etc.
\end{itemize}

\begin{figure*}
\centering
\includegraphics[angle=0, width=0.495\textwidth]{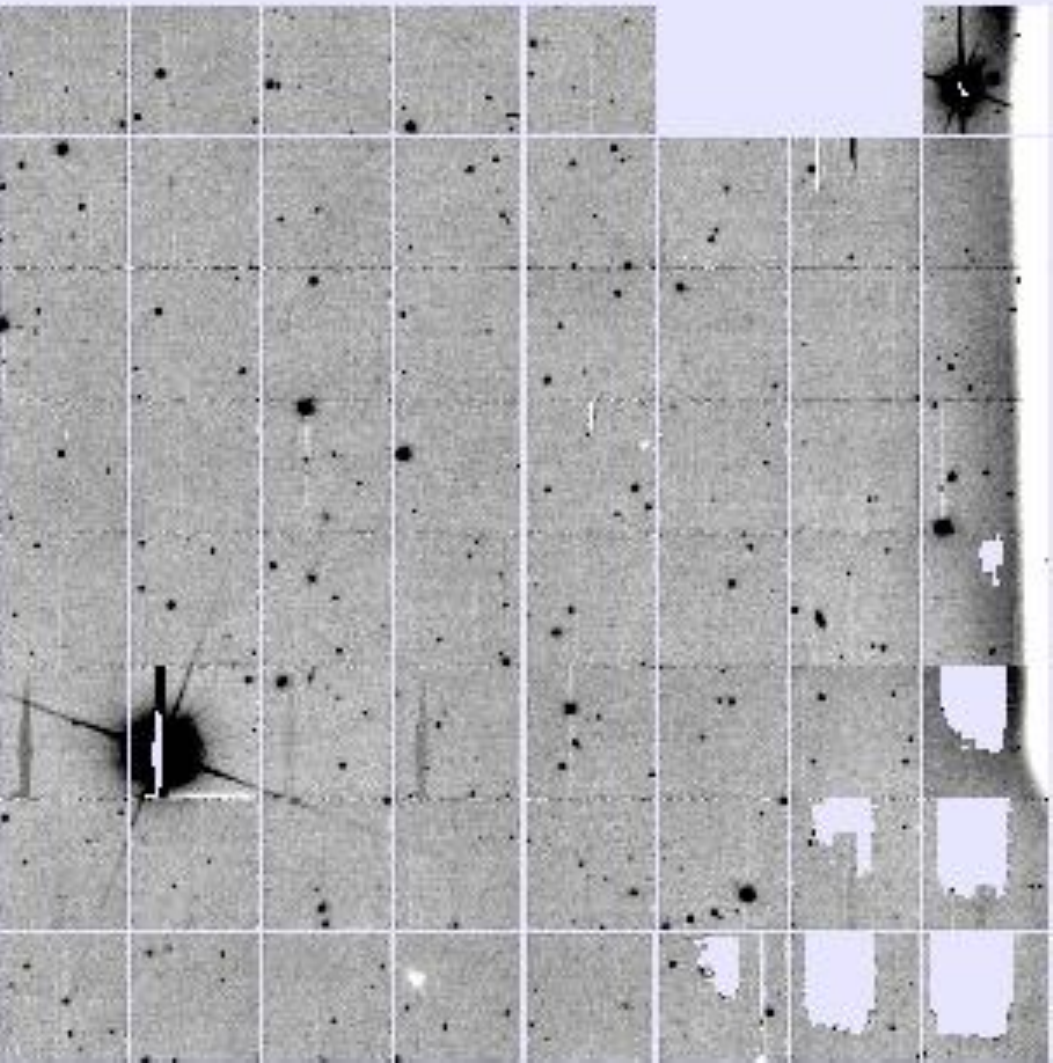}
\includegraphics[angle=0, width=0.495\textwidth]{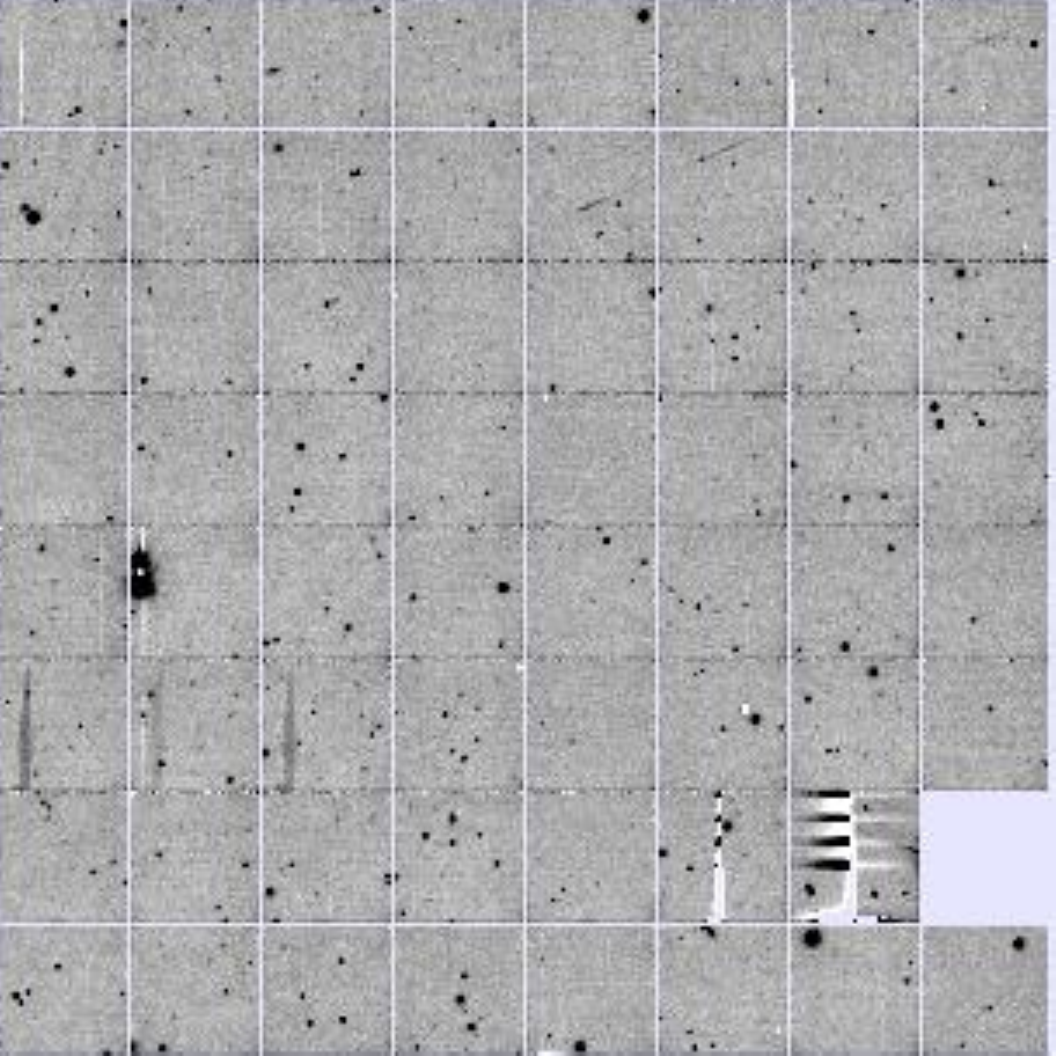}
\caption{Binned images of exposure o9272g0073o, showing a bright star on OTA04 and cell xy12 (left) which produces crosstalk on adjacent cellx xy02,xy22 and xy32. The right panel shows OTA24 of the same image, in which the bright star from the left panel produces crosstalk on cells xy02,xy12 and xy22.} \label{fig:crosstalk_example}
\end{figure*}

Given the way crosstalk arises, the dominant channels of crosstalk are expected to between OTAs and cells which are read out at the same time. For GPC1 this implies that we expect the major sources of crosstalk to be between cells located within the same row within an OTA. The second expected pattern is crosstalk between alternate OTAs (an offset of 2 columns in the focal plane), either on the same cell or in a cell within the same row. The third set of crosstalks expected are those between the OTAs located on the same STARGRASP controller (e.g. one column to the right). A look at an image with a bright star on one of the cells (Fig.~\ref{fig:crosstalk_example}) shows that these expected crosstalk channels are indeed present within the camera. The bright star located on cell xy12 of OTA XY04 leads to crosstalk on three of the neighboring cells within the same row (xy02,xy22,xy32) and crosstalk on 3 cells (xy02,xy12,xy22) on OTA XY24, which is the alternate device on the flex print assembly.

The example in Fig.~\ref{fig:crosstalk_example} shows that the crosstalk from bright stars can be identified fairly easily. However, for crosstallk channels with a high attenuation factor (meaning fainter crosstalk ghosts), it will be difficult to distinguish the ghost from real objects and background noise. Bearing this in mind, instead of using images from normal observation to determine the crosstalk channels, we performed a dedicated campaign which aimed to put a bright star in each cell of each OTA during twilight. The bright star targets were taken from the Yale Bright Star Catalog \citep{Hoffleit1982} and observed with short (0.1s) exposure times during twilight. The use of bright stars will allow us to find very faint crosstalk ghosts; while  observing during twilight means the presence of interloper stars is kept to a minimum. The aim was to cover all cells on all OTAs covering all the 8$\times$8$\times$60=3840 combinations. Over the course of a year (2019-01-03 to 2020-02-03), a total of 3789 exposures were taken, covering most combinations. 
 
\subsection{Bright Star Analysis}\label{subsec:bright_star_analysis}
To identify crosstalk without being influenced by any masking or flagging currently set up within the IPP system, the crosstalk analysis was performed using images on which only detrend reduction was applied \citep{Waters2020}. The on-cell position of each bright star was determined as the average position of the saturated star cores , after which we can look for crosstalk signals at the same position on different cells and OTAs. The case of having multiple saturated stars within the same CCD cell was dealt with by checking for multiple concentrated clumps of saturated pixels, which were each analysed separately if present. We determined  aperture magnitudes (using inner/outer/BG radii = 40/60/100 pixels) for both the source star and the potential crosstalk ghost on each cell of each OTA of each image. To be able to filter out noise and artifacts, we also determine the signal-to-noise ratio and correlation coefficient between image cutouts (of size 100$\times$100 pixels) at each location. Combining all this leads to a dataset of roughly 14 million entries for which pixel data exists (some cells are dead or otherwise unusable), within which to look for crosstalk channels in GPC1.

\begin{figure*}
\centering
\includegraphics[angle=0, width=0.495\textwidth]{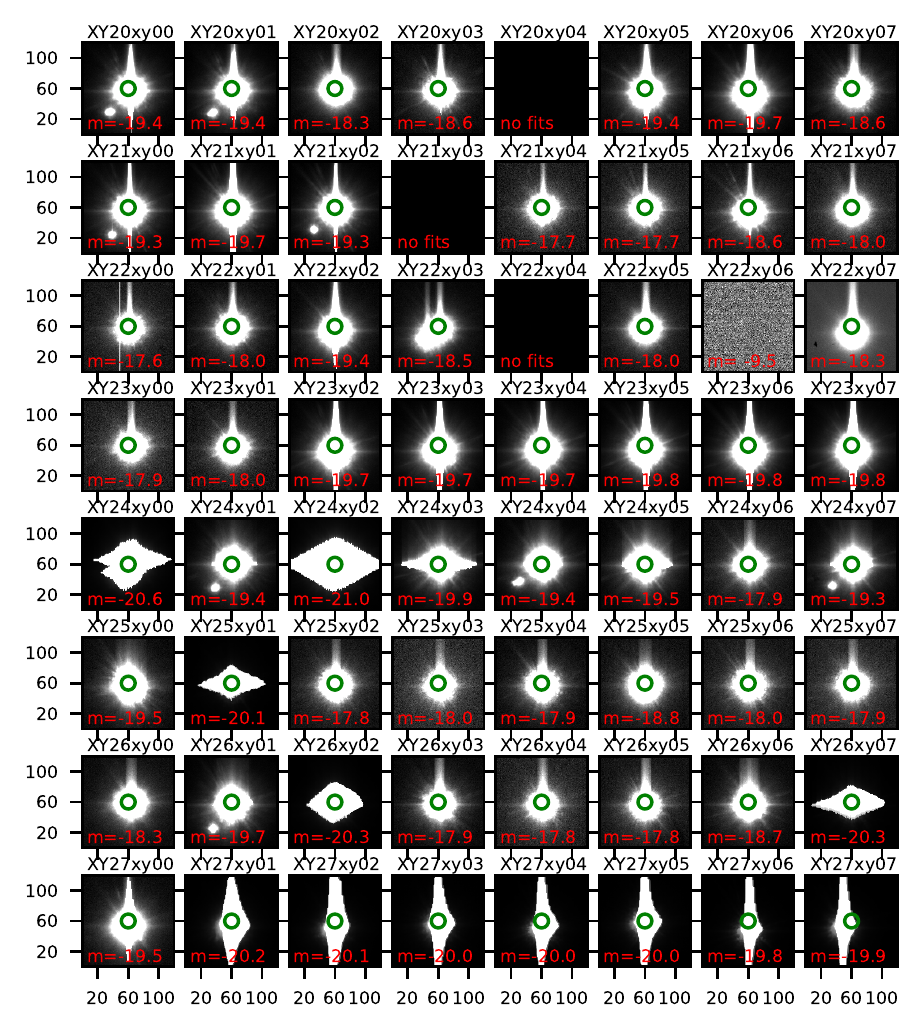}
\includegraphics[angle=0, width=0.495\textwidth]{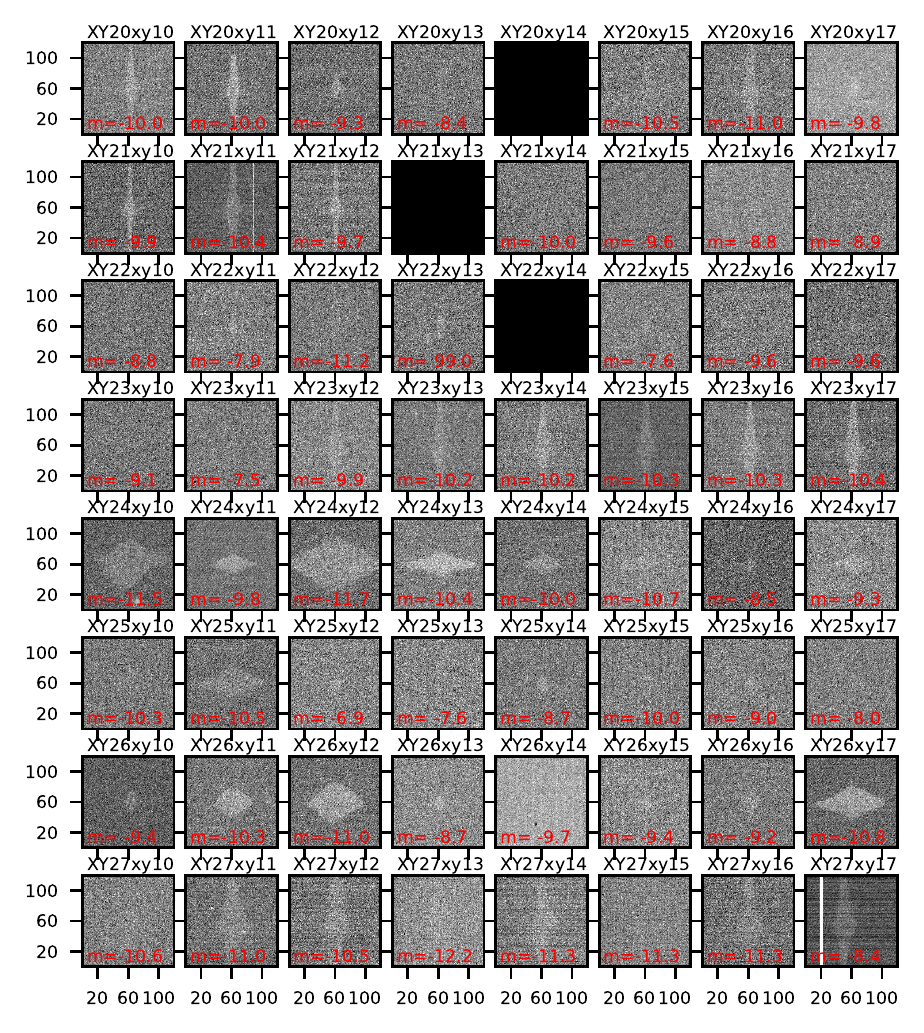}
\caption{An example of a particular crosstalk rule, as observed on all images within a particular column of OTAs and cells. In this case, stars on OTA columns 20-27 with cells xy00-xy07 (left panels) are producing crosstalk in images with cells xy10-xy17 of the same OTAs (right panels). Note that the footprint of the bright stars is preserved in the crosstalk, despite the vastly different aperture magnitudes of both images.} \label{fig:crosstalk_analysis}
\end{figure*}

Over the course of the analysis it became clear there were several instance in which the bright source star was not correctly placed within the target cell (or sometimes off entirely), leading to a bad detection of the crosstalk source. To eliminate these entries from our master table, we enforce a matching of brightness between the source aperture magnitude and the V-band entry in the Yale Bright Star Catalog, allowing a difference of 3~mags to account for different filters, noise and incomplete capture of the star within the aperture of the brightest targets.

\begin{figure*}
\centering
\includegraphics[angle=0, height=0.495\textwidth]{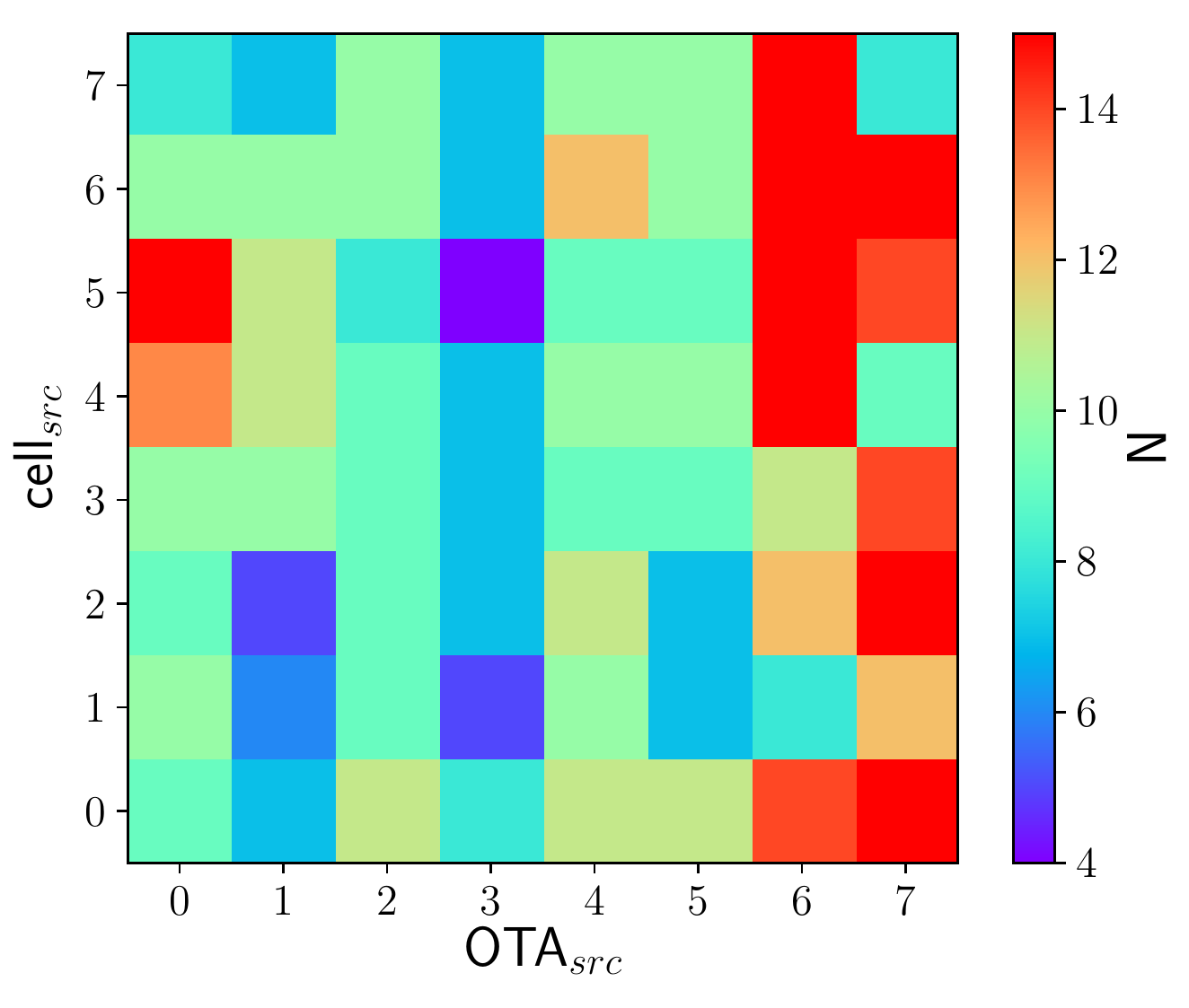}
\includegraphics[angle=0, height=0.495\textwidth]{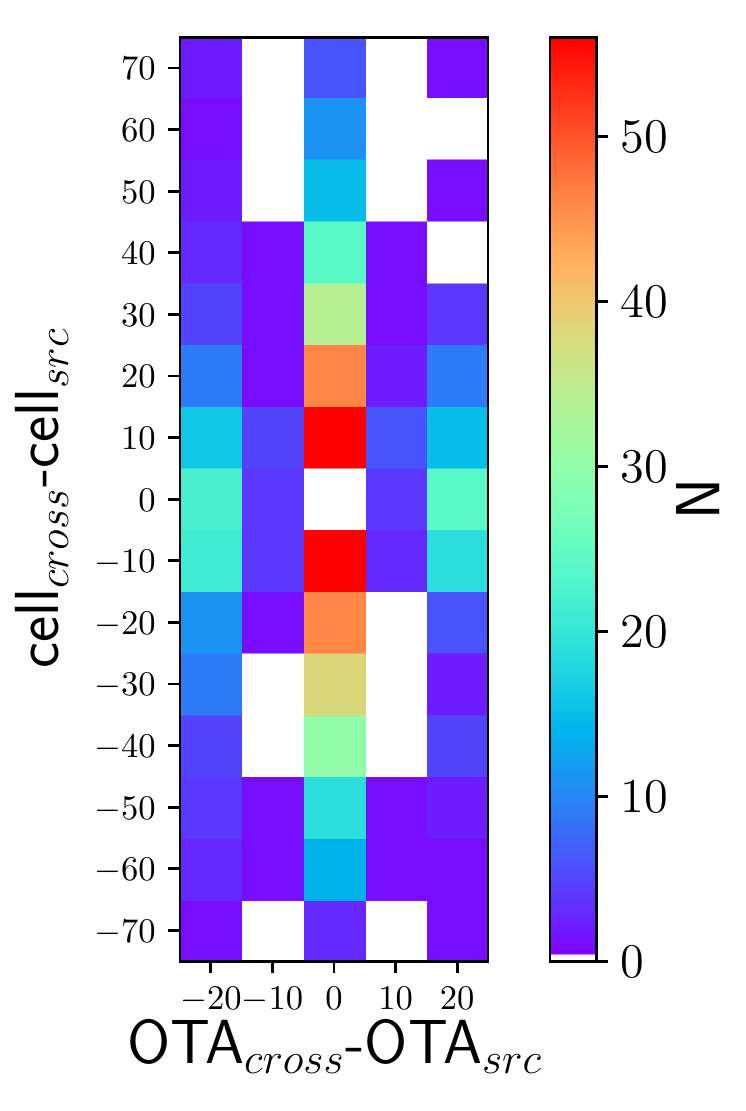}
\caption{Two dimensional histograms showing the number of crosstalk rules found for each set of OTA and cell columns within the focal plane, on the left. The right panel shows the offset in both the OTA and cell direction (destination - source) of the crosstalk rules, and the number found for each combination. The crosstalk rules apply to the entire column of chips and cells governed by the base coordinates (i.e. OTA$_{src}$=4 implies all OTAs within the fourth column (XY40-47)).} \label{fig:chip_cell_hist}
\end{figure*}

The master table was then filtered for clear noise or background areas by ensuring each crosstalk candidate has a signal to noise ratio greater than 1 and a correlation coefficient greater than 0.01. This left roughly 500,000 entries which might be viable crosstalk ghosts. To analyze this subset for crosstalk channels, we consider that a crosstalk channel will follow certain rules. Given the cell readout pattern, the same crosstalk offset and attenuation will apply to all cells within the same column within an OTA. Similarly, given the hardware layout, we can reasonably assume the same thing applies to all OTAs within the same column. This then allows us to look for clusters within the dataset corresponding to the same offset in OTA and cell between source and crosstalk ghost, belonging to the same overarching rule. 

With this proviso in place, we perform a clustering search of the crosstalk candidates to find those crosstalk channels for which at least half of the possible occupied cells result in a crosstalk ghost detection. An example of such a channel is shown in Fig.~\ref{fig:crosstalk_analysis} for one of the brighter channels. This figure shows that stars on OTA columns XY20-27 with cells xy00-xy07 are producing crosstalk in images with cells xy10-xy17 within the same OTA. This can be described as a channel with rule OTA2yXY0v to OTA2yXY1v (where both y and v run from 0 to 7) with an offset in OTAs of 0 and offset in cells of 10 (an octal movement of one column). For this rule, the difference in aperture magnitude between source star and crosstalk detection is 9.11$\pm$0.48. Note that the crosstalk is readily apparent in Fig.~\ref{fig:crosstalk_analysis}, with the footprints of the source stars clearly preserved despite the 9 mags of difference in aperture magnitude.

The clustering search resulted in a total of 181 crosstalk channels, all of which were examined by eye. The source positions and offset of all of these are consistent with the expected rules listed in section \ref{sec:method}, and we find no evidence of crosstalk occurring between OTAs which were not read out at the same time or not connected by the same flex print. Furthermore, as expected, the two halves of the focal plane are completely separate, with no channel crossing over the focal plane divide (despite being allowed by the simultaneity of the readout). 

Guided by the patterns inherent within the uncovered channels, and the observation that looking at the possible channel by eye allowed the authors to identify crosstalk ghosts which were classified as noise by the algorithm, we sought to improve the base analysis. To that end, plots similar to Fig.~\ref{fig:crosstalk_analysis} were generated for all the 1792 possible crosstalk channels resulting from the rules listed in section \ref{sec:method}. All of these images were then looked at by eye to determine if the channel showed crosstalk. The result of this search was a grand total of 640 channels with visible crosstalk, often allowing us to probe fainter crosstalk signals than otherwise possible.

\section{RESULTS}\label{sec:results}
With the set of crosstalk channels for GPC1 in place, we can look at the characteristics of the rules and their distribution across the focal plane. First off, we look at the number of crosstalk rules as a function of the OTA and cell column of their source star. The left panel of Fig.~\ref{fig:chip_cell_hist} shows that we find at least 4 crosstalk rules for each possible combination of OTA and cell columns, with a maximum of 14 rules mostly contained to the rightmost two columns of the focal plane. The third column of OTAs has the least number of rules with 52 total.
 
The right panel of Fig. \ref{fig:chip_cell_hist} shows the offset in both the OTA and cell direction of the crosstalk rules. As mentioned before, all rules are consistent with the expected behavior mentioned in section \ref{sec:method}. The majority (398 out of 640) of the rules applies to the inter-OTA crosstalk channels, with most rules showing only small offsets (1-2 columns) across the row of cells the source star is on. However, there are several cases in which the crosstalk movement extends all the way to the opposite edge of the row of cells from the source star.
 
Besides the inter-OTA crosstalk, there are also 38 rules in which the crosstalk is found one column to the left or right, and 204 rules where the crosstalk is offset by two columns. The channels offset by two columns cover most of the possible cell column offsets, while the channels offset by one column are more constrained to small cell offsets. Not every adjacent combination is covered, meaning that for some source OTA and cells we do find a rule for a cell offset of e.g. 1,2 and 4 columns, but not a rule for the offset of 3 columns. Therefore, it is possible there are more crosstalk rules present, but too faint for us to currently detect.

\begin{figure}
\centering
\includegraphics[angle=0, width=0.495\textwidth]{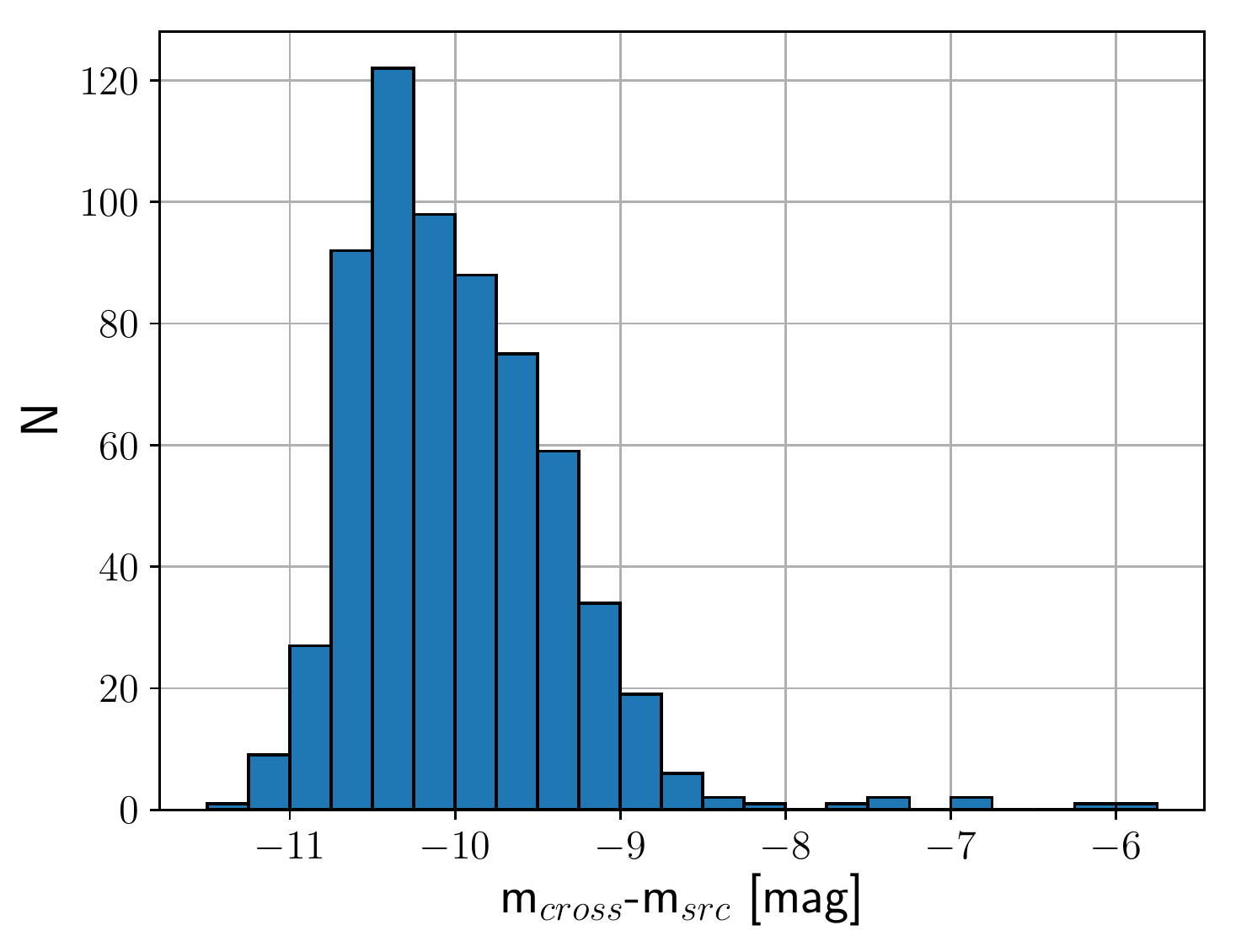}
\caption{The offset in magnitudes between the source star and crosstalk detection from our set of crosstalk rules. The histogram displays the number of rules as a function of magnitude, peaking at roughly m$_{cross}$-m$_{source}$=-10.25 mag. } \label{fig:crosstalk_magoffset_hist}
\end{figure}

An important property of the crosstalk channels is the difference between the brightness of the source object and the crosstalk ghost. To that end, we can compare the aperture magnitude of the source star and crosstalk ghost measurement of our set of rules and see the attenuation factor. Fig.~\ref{fig:crosstalk_magoffset_hist} displays a histogram of the magnitude offset for the full set of rules. The GPC1 crosstalk channels span a range of magnitude offsets, with the brightest rule leading to an offset of roughly 6 mags and the faintest rule offset by nearly 11 mags. In terms of flux attenuation factor, the bright rule attenuates by a factor of $\approx$250 while the faint rules attenuate flux by a factor $\approx$2.5$\times$10$^{4}$.

The bulk of the crosstalk rules are found around m$_{cross}$-m$_{src}\approx$-10.25, or a flux attenuating factor of 1.25$\times$10$^{4}$. The crosstalk ghosts resulting from these rules are faint, with an average signal to noise ratio of 0.64$\pm$0.1 within that bin. Therefore, these rules are very hard to identify from images obtained from normal observing modes. Furthermore, the crosstalk channels will not cause much interference during normal operations for identifying moving objects, with crosstalk ghost from the bright stars in typical fields being below the faint magnitude cut-off of the images. Nevertheless, in cases were a bright star with instrumental magnitudes close to -19 are observed, these will lead to a lot of crosstalk ghosts depending on its location within the focal plane.

% Refs on stellar density
% http://articles.adsabs.harvard.edu//full/1981ApJ...246..122B/0000123.000.html (see also Allen in this)
% https://ui.adsabs.harvard.edu/abs/2013AN....334..823Z/abstract
% https://ui.adsabs.harvard.edu/abs/2013AstBu..68..481Z/abstract

\begin{figure}
\centering
\includegraphics[angle=0, width=0.495\textwidth]{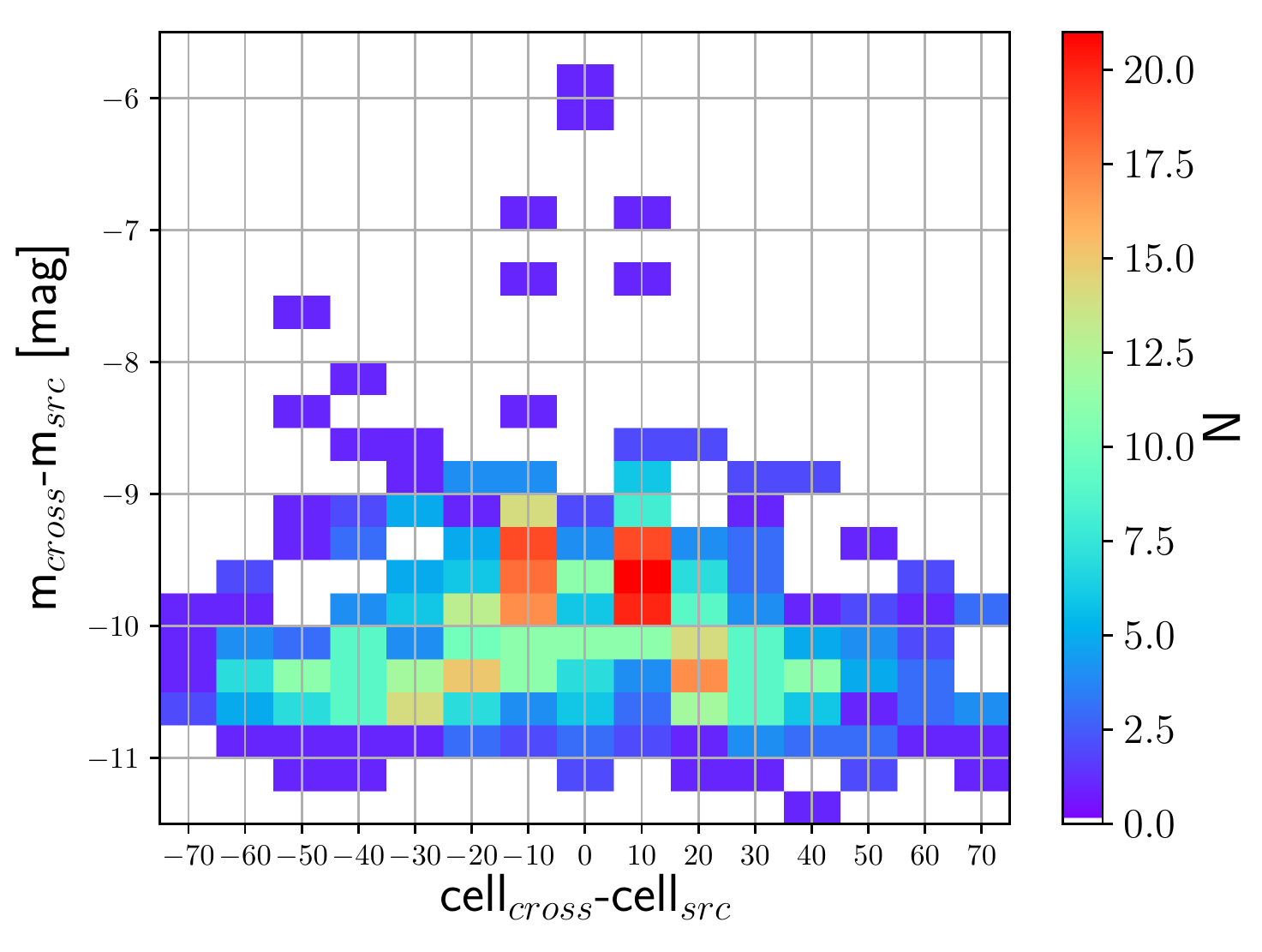}
\includegraphics[angle=0, width=0.495\textwidth]{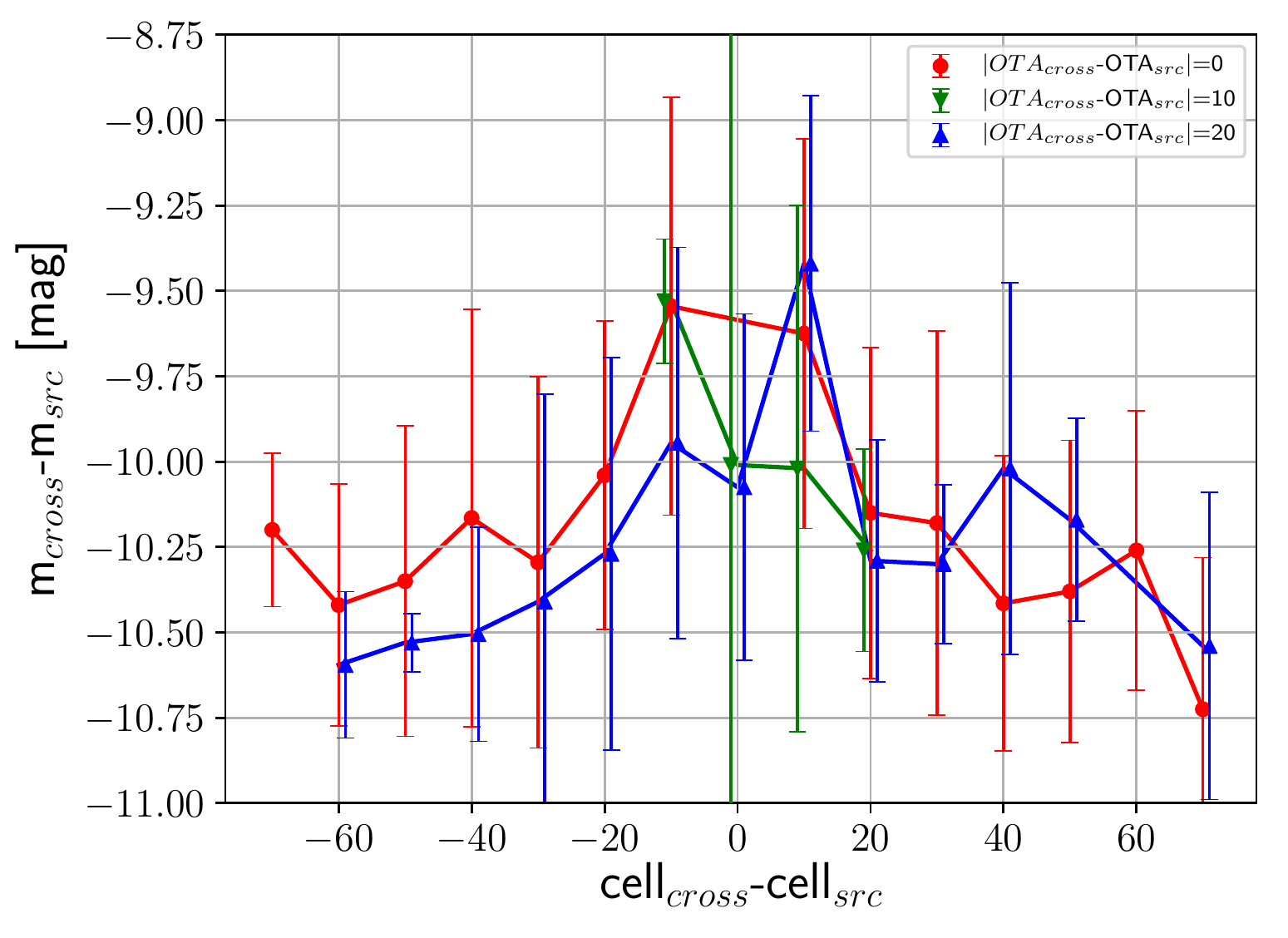}
\caption{The offset in magnitudes from our set of crosstalk rules, as a function of the offset between source and crosstalk in cells within the same row. The top panel shows the magnitude offsets, while the bottom panel shows a binned representation of the distribution as medians and standard deviations per bin for each OTA offset channel (with errorbars offset for visibility).} \label{fig:crosstalk_magoffset}
\end{figure}

Besides the general census of crosstalk attenuation factors, it is worth looking at the magnitude offsets as a function of crosstalk movement. There is no clear difference as a function of chip column movement in the dataset, with offsets consistent within the magnitude uncertainties. However, there is a difference in magnitude offset as a function of the cell column movement, as shown in the top panel of Fig.~\ref{fig:crosstalk_magoffset}. The distribution of magnitude offsets forms a rough pyramid shape, with a structure clearly mirrored around zero. The rules with greater cell column offsets display a greater magnitude offset on average. This lines up with the idea that the crosstalk flux is more attenuated the more cell connection lines are traversed. 

To analyze this in some more detail, the bottom panel of Fig. \ref{fig:crosstalk_magoffset} displays the median magnitude offset in each bin of cell column offset, and split out in the three possible OTA offset channels. It is readily apparent that the median values decrease for greater offset in cells, although the standard deviations are large, owing to the faintness of the crosstalk ghosts. It also appears that the crosstalk ghosts for crosstalk rules offset in OTA are in general fainter than those constrained to their source OTA.

Standing out from the bulk of the crosstalk rules are the bright points with relatively small magnitude offsets (e.g. m$_{cross}$-m$_{src}$>-8). A closer look at these rules reveals that they are typically found mirrored across the focal plane, with a similar offset on both rules. An example of some of the mirrored sets are:
\begin{itemize}
    \item OTA2yXY3v to OTA3yXY3v with mag offset $-5.98\pm$0.30 and OTA5yXY3v to OTA4yXY3v with mag offset $-6.1\pm$0.48 
    \item OTA2yXY5v to OTA2yXY6v with mag offset $-6.83\pm$0.25 and OTA5yXY5v to OTA5yXY6v with mag offset $-7.37\pm$0.86
    \item OTA2yXY6v to OTA2yXY5v with mag offset $-6.86\pm$0.23 and OTA5yXY6v to OTA5yXY5v with mag offset $-7.46\pm$0.93
    \item OTA2yXY7v to OTA3yXY2v with mag offset $-7.71\pm$0.68 and OTA5yXY7v to OTA4yXY2v with mag offset $-8.39\pm$0.62
\end{itemize}
Besides being bright, these channels are also relatively isolated, in the sense that they do not appear to 'spread out' along cell readout lines (e.g., OTA2yXY3v moving to both OTA2yXY1v, OTA2yXY2v and OTA2yXY4v), unlike the other rules. Given their brightness, these adjacent rules would have easily shown up, which is not the case. It seems likely that a different type of crosstalk is producing these sets of rules, which are in place on both sides of the focal plane.

\section{CONCLUSIONS}\label{sec:conclusions}
In this work, we have utilized data from bright stars to characterize crosstalk within the Pan-STARRS GPC1 camera. While the current flagging of crosstalk within IPP was good enough to eliminate bright ghosts, our analysis shows that a large number of faint ghosts can also be present and likely led to spurious signals that could be mistaken for faint moving objects if it wasn't for the manual inspection of candidates. With the new rules in place, the number of false positives being eye-balled by the MOPS team should go down significantly in crowded fields.

This analysis has shown that a dedicated campaign of bright star observations is a powerful tool for the study of crosstalk channels, and worth considering on new facilities such as LSST. That being said, it represents a sizeable investment of time for cameras with focal planes made up of many CCDs. The process of covering all CCDs with a bright star could potentially be sped up by targeting appropriate open clusters of bright stars such as the Pleiades, and covering multiple CCD combinations at once.

The LSST team has attempted to characterize crosstalk effects in a similar manner as described here, through the use of both a spot projector and analyzing many images with a suitable number of stars \citep{Snyder2021}. The LSST analysis is limited to inter-amplifier crosstalk, which makes up the bulk of our own crosstalk signals as well. It would be interesting to see if any crosstalk signals are detected that cross over the amplifier boundaries.

The grouping of crosstalk signals into groups of 8$\times$8 columns in OTAs and cells provided us with a powerful way of identifying coordinated crosstalk movement and enough sample size to obtain attenuation statistics. This grouping is supported by the hardware and focal plane layout of the GPC1 camera \citep{Onaka2008, Onaka2012}. The coordinated movement is readily visible in Fig.~\ref{fig:crosstalk_analysis}, which shows that stars with different saturated core morphology on adjacent cell columns end up next to each other, following bulk movement.

Fig.~\ref{fig:crosstalk_analysis} also makes it clear that the footprint of the crosstalk ghost signal is the same as that of the source star, despite a significant attenuation of the source flux. This is an important characteristic to consider when flagging or masking the ghosts since the size of the mask should be based on the source brightness and not the ghost brightness. In the current iteration of masking within IPP, the crosstalk features are masked using a circular mask only.

The resulting set of crosstalk rules presented in section \ref{sec:results} covers all of the GPC1 OTA and cell combinations, meaning each CCD leads to crosstalk at some level. Most of the crosstalk occurs within the same OTA device, across rows of cells, which lines up with what we expect. Charge spreads out across the cell readout lines, attenuating more and more the further it travels. Fig.~\ref{fig:crosstalk_magoffset} shows the magnitude offset of each rule as a function of cell column offset, clearly showing the mirrored nature of the rules and the increasing attenuation further away from the source star position.

We find no evidence within our dataset of crosstalk movement that does follow the possible pathways laid out by the hardware layout. The crosstalk movement to other OTAs within the same hardware flex-print assembly also makes sense given the camera hardware, and confirms the assumption that crosstalk within the GPC1 system is restricted to individual flex-print assemblies only. 

Our analysis of the crosstalk attenuation factor shows a great variety of different crosstalk responses. The brightest rule attenuates flux by a factor of 2.5$\times$10$^{2}$ while the faintest rule has a factor 2.5$\times$10$^{4}$. The most frequent crosstalk rule has m$_{cross}$-m$_{src}\approx$-10.25, which corresponds to a flux attenuating factor of 1.25$\times$10$^{4}$. The fact that these rules have an average signal to noise ratio of 0.64$\pm$0.1 means they are hard to find by analyzing images from normal scientific operations.

The distribution of magnitude offsets as a function of crosstalk movement shown in Fig.~\ref{fig:crosstalk_magoffset} reveals a clear dependence on cell column offsets consistent with the idea that the source star charge is attenuated during the traversal of cell readout lines. While we can see the trends, the uncertainties on aperture magnitude measurements are large at this stage. This is something that deserves to be investigated further, as it can be used for a more general characterization of the crosstalk attenuation without invoking  specific rules.

Moving forward, a similar analysis will be conducted on the second Pan-STARRS telescope, to characterize crosstalk within the GPC2 camera. The hardware setup of both telescopes is similar enough that we expect similar behaviors to occur there, but potentially at different levels.

\acknowledgments
The Pan-STARRS1 Surveys (PS1) have been made possible through
contributions of the Institute for Astronomy, the University of
Hawaii, the Pan-STARRS Project Office, the Max-Planck Society and its
participating institutes, the Max Planck Institute for Astronomy,
Heidelberg and the Max Planck Institute for Extraterrestrial Physics,
Garching, The Johns Hopkins University, Durham University, the
University of Edinburgh, Queen's University Belfast, the
Harvard-Smithsonian Center for Astrophysics, the Las Cumbres
Observatory Global Telescope Network Incorporated, the National
Central University of Taiwan, the Space Telescope Science Institute,
the National Aeronautics and Space Administration under Grant
No. NNX08AR22G issued through the Planetary Science Division of the
NASA Science Mission Directorate, the National Science Foundation
under Grant No. AST-1238877, the University of Maryland, and Eotvos
Lorand University (ELTE) and the Los Alamos National Laboratory.

\bibliographystyle{apj}
\bibliography{Bibliography}

\end{document}